%% file: root.tex
\documentclass[letterpaper, 10 pt, conference]{ieeeconf}  

\IEEEoverridecommandlockouts                              

\usepackage{graphicx} 

\usepackage{hyperref}
\usepackage{orcidlink}
\usepackage[acronym,ucmark,translate]{glossaries}
\setacronymstyle{long-short}
\glsdisablehyper
\usepackage{siunitx}

\newcommand\copyrighttext{%
    \footnotesize \textcopyright 2023 IEEE. Personal use of this material is permitted. Permission from IEEE must be obtained for all other uses, in any current or future media, including reprinting/republishing this material for advertising or promotional purposes, creating new collective works, for resale or redistribution to servers or lists, or reuse of any copyrighted component of this work in other works.
    DOI: \href{https://doi.org/10.1109/IROS55552.2023.10341256}{10.1109/IROS55552.2023.10341256}}
\newcommand\copyrightnotice{%
\begin{tikzpicture}[remember picture,overlay]
\node[anchor=south,yshift=10pt] at (current page.south) {\fbox{\parbox{\dimexpr\textwidth-\fboxsep-\fboxrule\relax}{\copyrighttext}}};
\end{tikzpicture}%
}

\title{\LARGE \bf
Design Space Exploration on Efficient and Accurate Human Pose Estimation from Sparse IMU-Sensing
}

\author{Iris F\"urst-Walter \orcidlink{0000-0001-8454-634X}, Antonio Nappi, Tanja Harbaum \orcidlink{0000-0001-7310-567X}, J\"urgen Becker \orcidlink{0000-0002-5082-5487}
\thanks{This work has been supported by the project “Stay young with robots” (JuBot). The JuBot project was made possible by funding from the Carl-Zeiss-Foundation.}
\thanks{The authors are with the Institut fuer Technik der Informationsverarbeitung, Karlsruhe Institute of Technology, Germany. {\tt\small \{fuerst, harbaum, becker\}@kit.edu}}%
\thanks{Code available online at \textbf{https://github.com/itiv-kit/dse-sparse-imu}.}%
}

\begin{document}

\maketitle
\copyrightnotice
\thispagestyle{empty}
\pagestyle{empty}

\begin{abstract}
Human Pose Estimation (HPE) to assess human motion in sports, rehabilitation or work safety requires accurate sensing without compromising the sensitive underlying personal data.
Therefore, local processing is necessary and the limited energy budget in such systems can be addressed by Inertial Measurement Units (IMU) instead of common camera sensing.
The central trade-off between accuracy and efficient use of hardware resources is rarely discussed in research.
We address this trade-off by a simulative Design Space Exploration (DSE) of a varying quantity and positioning of IMU-sensors.
First, we generate IMU-data from a publicly available body model dataset for different sensor configurations and train a deep learning model with this data.
Additionally, we propose a combined metric to assess the accuracy-resource trade-off.
We used the DSE as a tool to evaluate sensor configurations and identify beneficial ones for a specific use case.
Exemplary, for a system with equal importance of accuracy and resources, we identify an optimal sensor configuration of 4 sensors with a mesh error of 6.03 cm, increasing the accuracy by~\SI{32.7}{\percent} and reducing the hardware effort by two sensors compared to state of the art.
Our work can be used to design health applications with well-suited sensor positioning and attention to data privacy and resource-awareness.

\end{abstract}

\input{acronyms}
\input{sections/1_introduction.tex}
\input{sections/2_related_work.tex}
\input{sections/3_methodology.tex}
\input{sections/4_results.tex}
\input{sections/5_conclusion.tex}

\addtolength{\textheight}{-12cm}   







\bibliographystyle{IEEEtran}
\bibliography{root}

\end{document}

%% file: acronyms.tex
\newacronym{ai}{AI}{Artificial Intelligence}
\newacronym{dse}{DSE}{Design Space Exploration}
\newacronym{smpl}{SMPL}{Skinned Multi-Person Linear Model}
\newacronym{imu}{IMU}{Inertial Measurement Unit}
\newacronym{hpe}{HPE}{Human Pose Estimation}
\newacronym{dnn}{DNN}{Deep Neural Network}
\newacronym{sip}{SIP}{Sparse Inertial Poser}
\newacronym{dip}{DIP}{Deep Inertial Poser}
\newacronym{pip}{PIP}{Physical Inertial Poser}
\newacronym{rnn}{RNN}{Recurrent Neural Network}
\newacronym{lstm}{LSTM}{Long Short-Term Memory}
\newacronym{mmm}{MMM}{Master Motor Map}
\newacronym{mocap}{mocap}{motion capture}

%% file: sections/1_introduction.tex
\section{Introduction}
Assessment of human motion enriches many applications in sports, rehabilitation or work safety to monitor the movement quality and correctness~\cite{Avola2019}.
These applications require both an accurate \gls{hpe} and sensitive processing of the underlying personal data.
Especially in rehabilitation applications, privacy must be carefully respected as the acceptance of continuous monitoring systems depends on data security.
To preserve personal information, data locality and local computing is preferred over processing on remote systems.
However, due to battery-powered operation, local computing systems have to operate on a strict energy budget, which can be addressed by partitioning the data processing on different compute nodes~\cite{Kress2022} or data-efficient \gls{mocap} with \glspl{imu} instead of common camera sensing~\cite{vanSchaik2020}.

\begin{figure}[tb]
    \centering
    \includegraphics[width=\columnwidth]{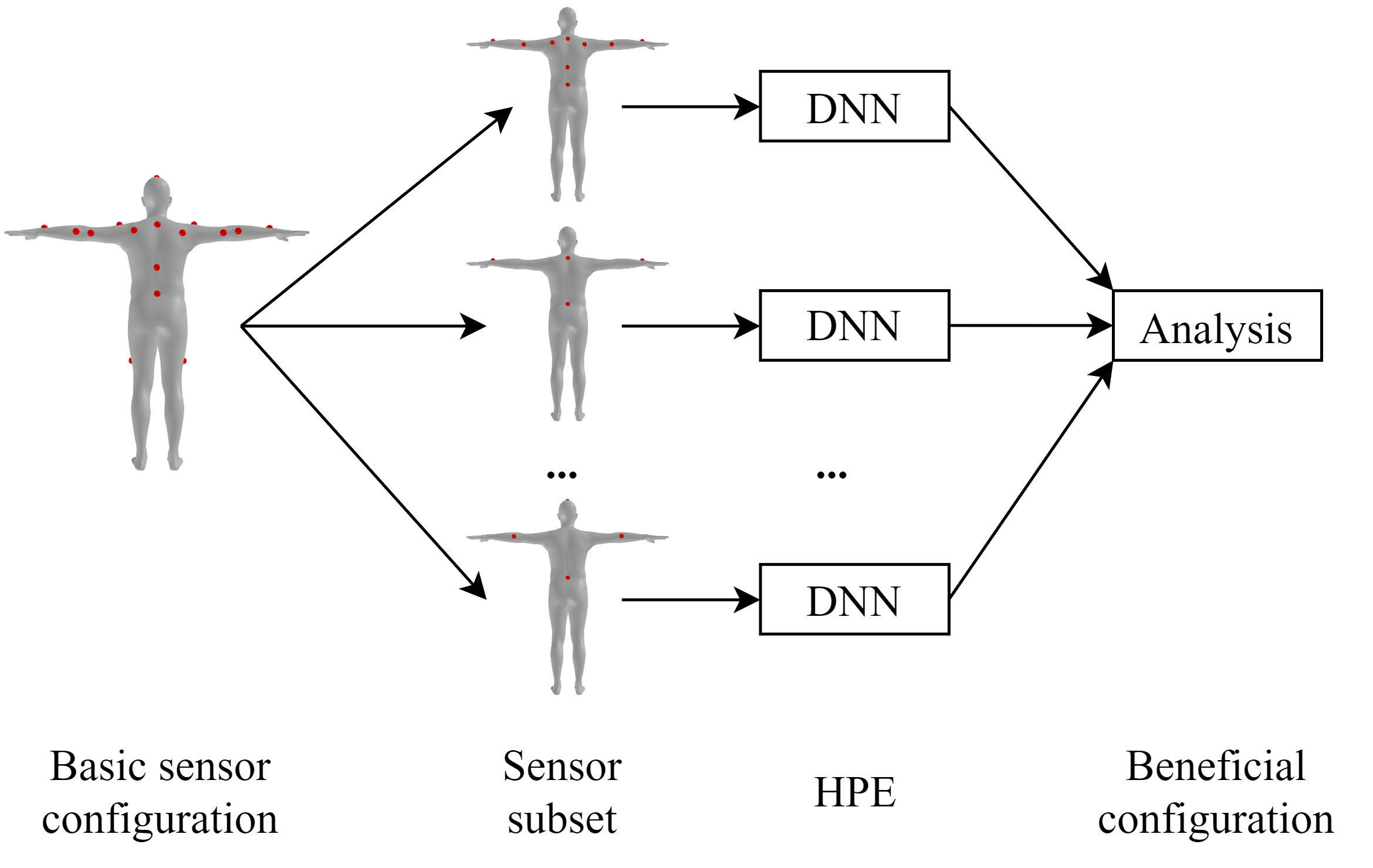}
    \caption{\textbf{Methodology of our Design Space Exploration.} We define a basic sensor configuration and synthesize IMU-data for sensor subsets. With this data, we train a deep neural network and evaluate diverse error metrics for \gls{hpe} of each subset. Finally, we analyse all experiments to identify a beneficial sensor positioning.}
    \label{fig:title_image}
\end{figure}

For \gls{imu}-based human motion tracking systems, the number and positioning of the sensors is crucial for an accurate \gls{hpe}.
The more sensors are attached to the human body, the more motion information is captured, but on the other hand, the hardware resource overhead, concerning e.g. energy consumption, available bandwidth or wiring, increases.
Therefore, a trade-off between accuracy and resources exists and a compromise for sparse \gls{imu}-sensing has to be found in each use case individually. 
To address this trade-off and establish a general methodology to evaluate systems with different sensor configurations, we perform a \gls{dse} on the number and positioning of \gls{imu}-sensors on the human body and propose a combined metric for assessment of the accuracy-resource trade-off.
This metric allows to weigh the importance of hardware resources in relation to prediction accuracy defined by the system designer in a specific use case.

Since mounting the sensors is time-intense and error-prone, we synthesize \gls{imu}-data for virtual sensors defined on the body model in \gls{mocap} datasets.
To rank different sensor configurations, we train a \gls{dnn} on the synthesized \gls{imu}-data and compare the resulting \glspl{hpe} with our combined metric.

We summarize our contribution as
\begin{itemize}
    \item Synthesis of \gls{imu}-data from a body model dataset with additional noise, including ground truth labeling for supervised learning of \gls{hpe}.
    \item Automated \gls{dse} of variable number and positioning of sensors on the human body with analysis of different sensor setups.
    \item Evaluation of beneficial sensor configuration in terms of the accuracy-resource trade-off of the \gls{hpe} using our introduced combined metric.
\end{itemize}

Real system design of body-mounted \gls{imu} motion tracking is time-consuming, where our \gls{dse} can be used as a tool  to identify beneficial positions and accelerate the design of fabric-integrated sensor systems.
We consider our investigation of sparse \gls{imu}-sensing to enable data-efficient and at the same time resource-aware health systems relying on an accurate \gls{hpe} to pave the way for appropriate application in movement monitoring and pose correction.

%% file: sections/2_related_work.tex
\section{Related Work}

\acrfull{hpe} from sparse \gls{imu}-sensing and the assessment of different sensor configurations relies on several areas spanning from the model estimating the \gls{hpe}, used learning data to positioning of the sensors.
Previous work in those areas will be presented in the following.

\subsection{Deep learning based Human Pose Estimation (HPE) from sparse IMU-sensing}
Different deep learning based models were investigated to estimate human pose from sparse \gls{imu}-data, enabling applications with lower mounting time and higher comfort due to fewer sensors.

\textbf{\gls{sip}}~\cite{Marcard2017} showed the potential of \gls{hpe} from a sparse \gls{imu}-configuration of six sensors to enable human \acrlong{mocap} in the wild.
The human pose is estimated offline in two stages.
The authors further introduced a new evaluation metric, the \gls{sip}-error, which only takes the orientation error of upper arms and thighs into consideration.
However, this approach is not suited for real-time application.

\textbf{\gls{dip}}~\cite{Huang2018} followed \gls{sip} with the main goal to improve applicability in online-systems.
Real-time capability is enabled by considering a limited number of past and future data and a \gls{rnn} is used to directly estimate body model parameters from orientation and acceleration inputs.
The authors further published a synthesis method to generate IMU-data from \gls{mocap} datasets.
Therefore, motions are mapped to a body model, on which virtual IMU sensors are defined.
For each virtual sensor, orientation and acceleration are calculated.

\textbf{TransPose}~\cite{Xinyu2021} introduced a two-stage approach to estimate pose as well as global translation of human motion.
Similar to \gls{dip}, bidirectional \glspl{rnn} are used with \gls{lstm} cells to estimate human pose.
The authors claim to be more accurate and computationally more efficient and the model achieves processing at \SI{90}{fps}.

\textbf{\gls{pip}}~\cite{Xinyu2022} was recently published and extends TransPose by a physical-aware model to estimate human pose, initial joint torques and ground reaction forces.
The term \textit{physical-aware} refers to the property of respecting physical constraints like jitter and ground penetrations.

\autoref{tab:sota_dl_hpe_evaluation} gives an overview of the currently available \gls{hpe}-systems.
It can be seen that \gls{pip} achieves the most accurate prediction, but it deployed the most complex estimator.

\subsection{Database} \label{sec:database}
In deep learning, the underlying database is crucial to train a reliable model.
Therefore, commonly used datasets for \gls{hpe} are presented.


\textbf{DIP-IMU}~\cite{Huang2018} recorded solely \gls{imu}-data captured by an Xsens motion tracking system with 17 sensor nodes.
The placement on the body is similar to Vicon, but not identical.
The recording volume is 92 minutes of 330,178 frames at \SI{60}{fps}.
Nine male and one female subject performed five different motions classes, including controlled arm and leg movements, locomotion, jumping jacks or boxing, and interactions with objects while sitting.

\textbf{AMASS}~\cite{Mahmood2019} is an actively developing collection of \gls{mocap} datasets, including \gls{smpl}-data of currently 24~datasets.
Overall, 500 subjects performed 17,916 motions, totaling a recording volume of 3,772.45 minutes at \SI{60}{fps}.
Collected motions include whole-body pose motions as well as detailed hand/finger exercises.

In addition to directly recording \gls{imu}-data from real movements, \cite{Huang2018} proposed synthesis from \gls{mocap} datasets.
Therefore, the authors placed virtual \gls{imu} sensors in AMASS and generated orientation and acceleration data for each sensor in each frame of the motion.

\begin{figure}[tb]
     \centering
     \begin{tabular}{ c }
         \includegraphics[scale=0.9]{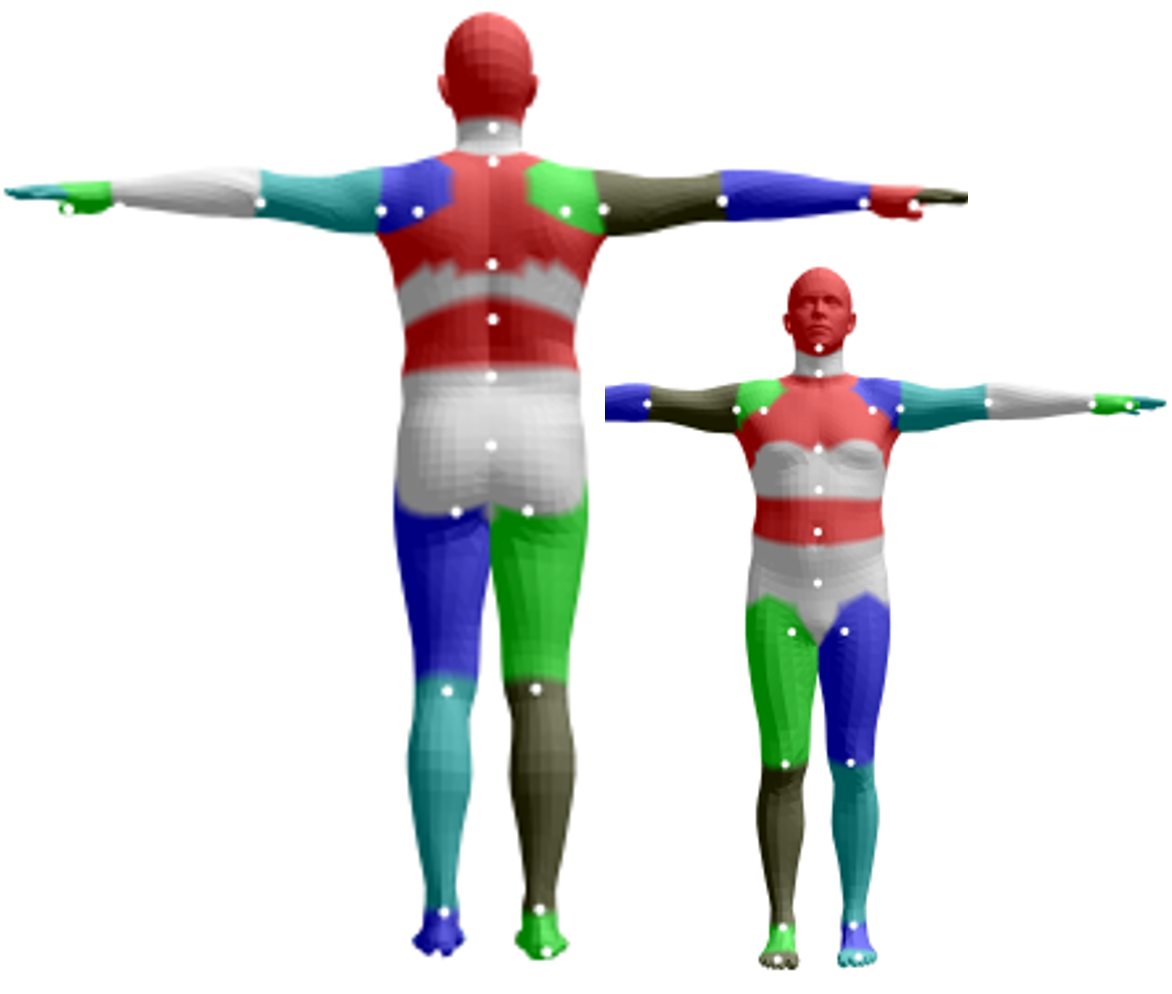} \\
         \small (a) Segmentation. Images from~\cite{Loper2015}
         \label{fig:SMPL_segments}\\
         \includegraphics[scale=0.6]{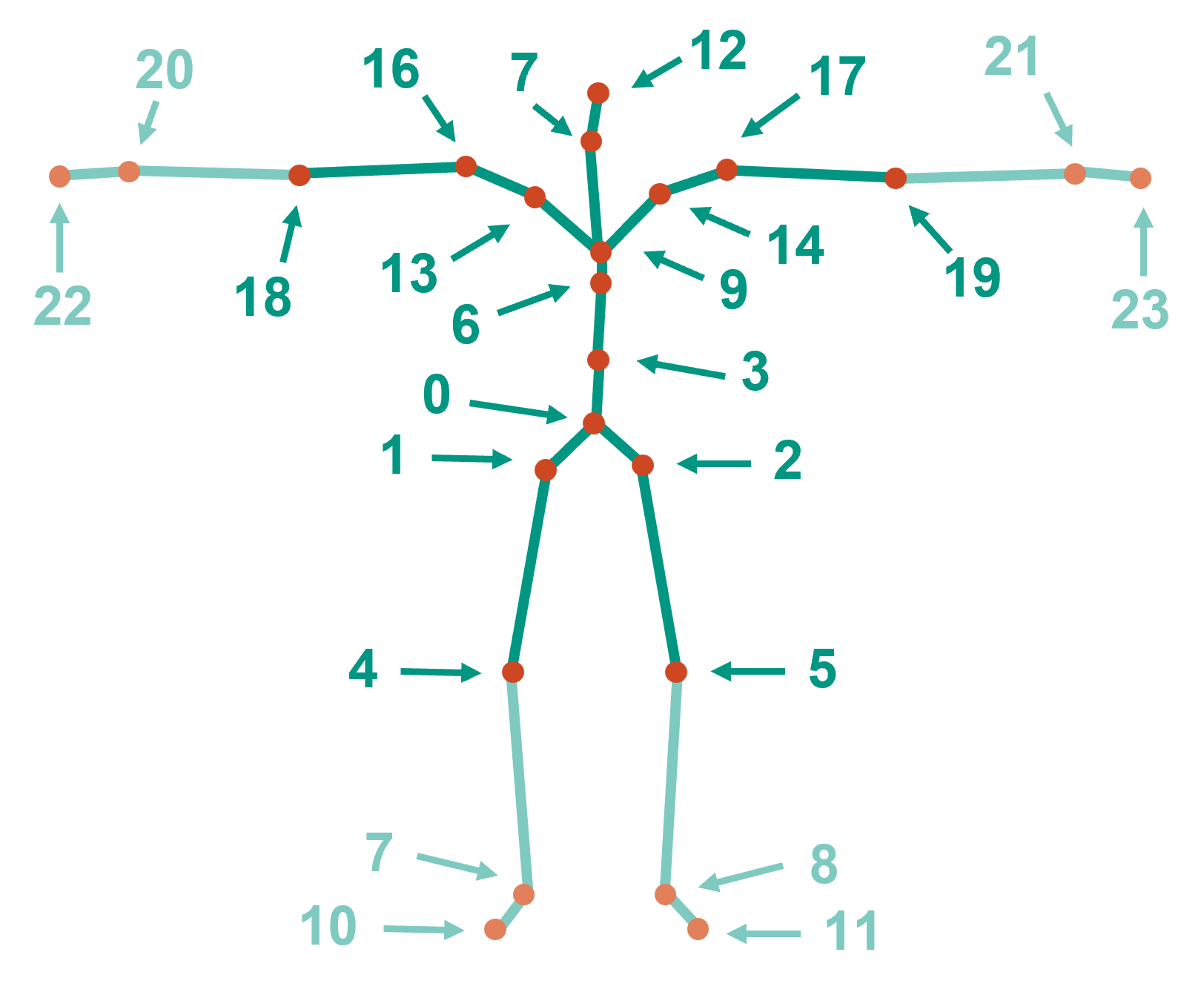} \\
         \small (b) Skeleton with Joint Indices. Inspired by \cite{Puchert2021}
         \label{fig:SMPL_joint_positions}\\
    \end{tabular}
    \caption{\textbf{Segments and joint positions of \gls{smpl}}~\cite{Loper2015}. (a) Segmentation of \gls{smpl}, where the white dots depict joints of the body model detailed in (b), shaded joints are not considered in \cite{Huang2018} and equally not in ours. (b) corresponds to the back view of the model, i.e., left side of (a)}
    \label{fig:SMPL}
\end{figure}

\textbf{\acrfull{smpl}}~\cite{Loper2015} is a common body model to generically represent human motion and is used in the AMASS dataset.
As shown in~\autoref{fig:SMPL}, \gls{smpl} simplifies the human body to a skeletal system of 24~segments with 23 joints and provides a wide range of body shapes.
The shape consists of a mesh of 6,890 vertices and deforms naturally with a given pose, mimicking soft-tissue dynamics.
For the synthesis of \gls{imu}-data, a virtual sensor can be placed at any vertex of the \gls{smpl} mesh.

\subsection{Sensor positioning}
\textbf{Xsens}~\footnote{https://www.xsens.com} is a commercial \gls{mocap} system with 17 wearable \glspl{imu}.
It provides a sensor per limb segment, shoulder, head, sternum and pelvis which is considered the root node.
This system has been used in, e.g. \gls{dip} to collect the DIP-IMU dataset, but the high amount of sensors requires high processing effort.

An optimal sensor positioning for instability detection was investigated by Steffan et al.~\cite{Steffan2017}.
They evaluated different IMU sensor configurations on the human body, including up to six sensors out of a basic configuration of 34 sensors, which was based on the optical marker placement of \gls{mmm}~\cite{Terlemez2014}.
For the \gls{dse}, they emulated \gls{imu}-data from the Whole-Body Human Motion Database~\cite{Mandery2015, Mandery2016} which is included in AMASS.
Depending on the F1-score for the instability detection, the best suited sensor configuration and best fitting model were identified.
The best instability detection was achieved with six sensors mounted on the right foot, wrists, left elbow, sternum and pelvis.

For motion classification of different displacement motions, Patzer et al.~\cite{Patzer2019} investigated a minimal sensor setup on an exoskeleton.
Instead of exhaustively testing every sensor configuration, they applied a wrapper-based method for three \glspl{imu} and seven force sensors.
Therefore, they started the \gls{dse} with evaluation of a single-sensor-system and defined the sensor with the most accurate motion classification as basic subset.
Iteratively, one new sensor was added to the basic setup and the new best subset was identified.
As a result, they identified a subset of six sensors to achieve the same accuracy as the basic configuration of ten sensors.

The latter \glspl{dse} both identify an optimal sensor configuration for their specific use case of instability detection and motion classification, but lack generalization for other applications and do not consider the hardware resources.

%% file: sections/3_methodology.tex
\section{Methodology to Evaluate Sensor Configurations}

\begin{figure}[tb]
    \centering
    \includegraphics[scale=0.9]{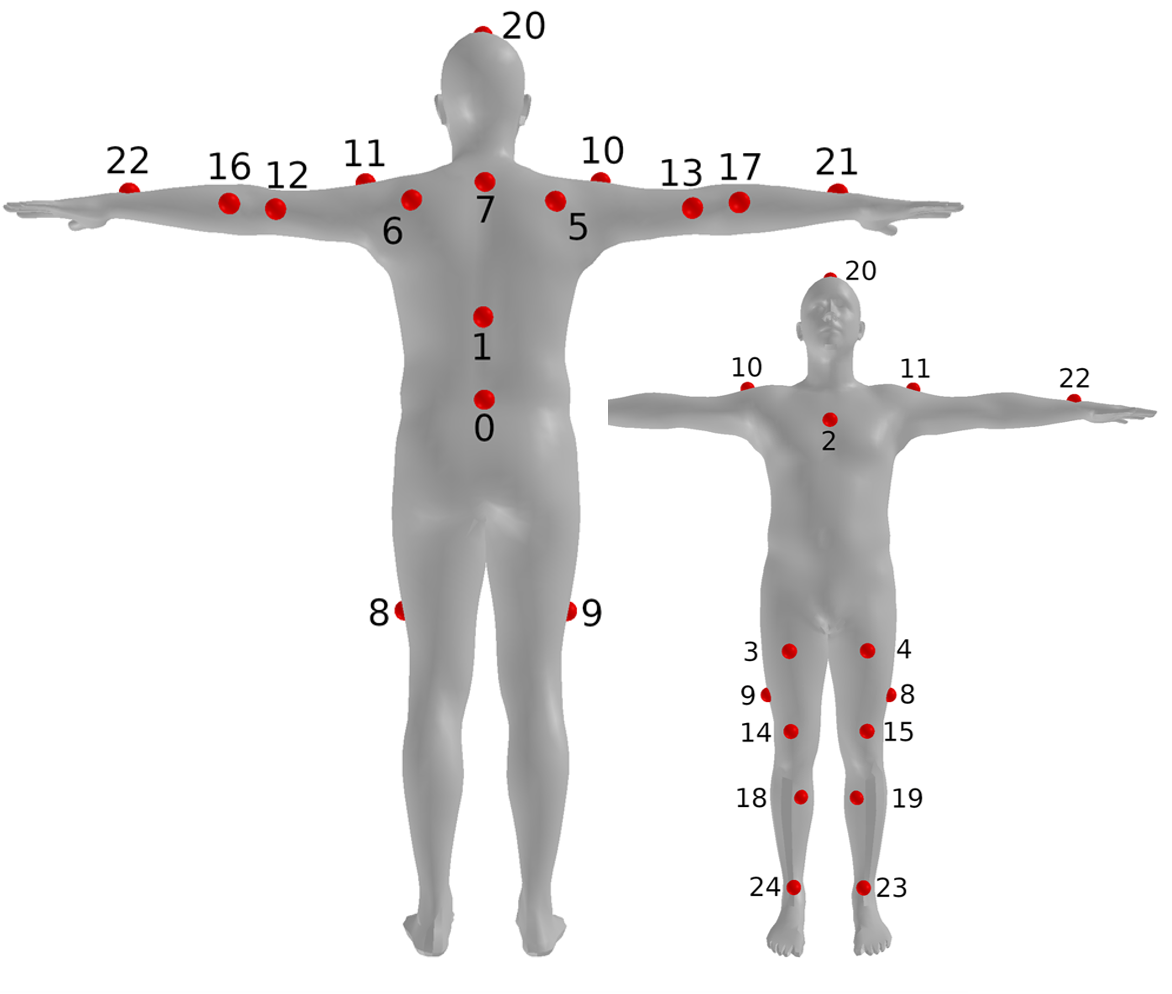}
    \caption{\textbf{Basic sensor configuration} illustrating all possible sensor positions considered for the \gls{dse}.}
    \label{fig:basic_sensor_configuration}
\end{figure}

We introduce a methodology to evaluate sensor configurations with a constrained \acrfull{dse} as a tool to support real system design of human \acrlong{mocap} from sparse \gls{imu}, and propose a combined metric to assess both accuracy and hardware resources.
For illustration of this methodology, we exemplary go through the design of a motion tracking system in rehabilitation to monitor full-body exercises of stroke patients~\cite{Avola2019}.
In this use case, we focus on monitoring the regain of symmetric performance of the exercises after half-sided paralysis and show how a suitable sensor configuration can be determined.

Therefore, we first define a basic sensor configuration as shown in \autoref{fig:basic_sensor_configuration}, where the sensor positions are inspired by the well-established configurations by Xsens and Steffan et al.~\cite{Steffan2017}.
For full-body capture, we add sensors around every limb joint and extend the sensor configuration to be symmetric, ensuring observability of symmetric execution of the rehabilitation exercises.
Our definition of the basic sensor configuration includes 25 sensors, i.e., five sensors per leg, four sensors per arm, one at each shoulder, three along the spine, one at the sternum, and one at the head.
The exact sensor positions on the \gls{smpl} model are depicted in~\autoref{fig:basic_sensor_configuration}.

To ensure reliable full-body recognition and feasibility in real systems, we restrict the maximum number of used sensors to 10 out of the 25 sensors.
Otherwise, this would lead to a combinatorial explosion of over 7 million sensor configurations.
To limit the considered sensor setups and therefore the computational effort of the \gls{dse}, we apply further constraints and reduce redundant caption of the segments by allowing only one sensor per segment, fixing the pelvis node with ID 0 as the root node and focusing on symmetric sensor configurations imposed by the use case.
These restrictions lead to a total of 2,249 configurations to evaluate.

For each sensor configuration, we synthesize \gls{imu}-data from a body-model dataset using the methods of \gls{dip}~\cite{Huang2018}.
For training, we use synthetic data from AMASS, for fine-tuning, synthetic data from DIP-IMU subjects 1 to 8, and for testing synthetic data from DIP-IMU subjects 9 and 10.
A holdout of five sequences from DIP-IMU subjects 1 to 8 is reserved as a validation dataset.
For sensor configurations that match the real IMU recordings in DIP-IMU, we proved feasibility of using synthetic data and validated the synthesis approach.
We further add white noise during synthesis to map statistical errors of real IMU-sensors and could therefore improve \gls{hpe} compared to synthesis of ideal sensors.

\begin{table}[tb]
    \centering
    \caption{\textbf{Error comparison} on DIP-IMU dataset. For equal importance of accuracy and resources, we identify a sensor setup with two sensors less than the DIP reference configuration and improve SIP by \SI{10.9}{\percent}, mesh by \SI{32.7}{\percent} and jitter by \SI{82.1}{\percent}, respectively.}
    \label{tab:sota_dl_hpe_evaluation}
    \begin{tabular}{c|c|c|c}
        \textbf{Method} & \textbf{SIP Err ($\si{deg}$)} & \textbf{Mesh Err ($\si{\cm}$)} & \textbf{Jitter ($\frac{\si{km}}{\si{s^3}}$)} \\ \hline
        SIP (offline)~\cite{Marcard2017} & 21.02$^{\mathrm{a}}$ & 7.71$^{\mathrm{a}}$ & 0.38$^{\mathrm{a}}$  \\
        DIP~\cite{Huang2018} & 17.10$^{\mathrm{b}}$ & 8.96$^{\mathrm{b}}$ & 3.01$^{\mathrm{a}}$ \\
        TransPose~\cite{Xinyu2021} & 16.68$^{\mathrm{b}}$ & 7.09$^{\mathrm{b}}$ & 1.46$^{\mathrm{b}}$ \\
        PIP~\cite{Xinyu2022} & 15.02$^{\mathrm{b}}$ & 5.95$^{\mathrm{b}}$ & $\mathbf{0.24}$$^{\mathrm{b}}$ \\ \hline
        DIP (ours, 6)$^{\mathrm{c}}$ & \textbf{13.63} & \textbf{5.89} & 0.49 \\
        \textit{DIP (ours, 4)}$^{\mathrm{d}}$ & \textit{15.24} & \textit{6.03} & \textit{0.54} \\
        \textit{DIP (ours, full)}$^{\mathrm{e}}$ & \textit{2.77} & \textit{1.91} & \textit{0.47} \\
        \multicolumn{4}{l}{$^{\mathrm{a}}$values are taken from~\cite{Xinyu2021}, $^{\mathrm{b}}$values are taken from~\cite{Xinyu2022}} \\
        \multicolumn{4}{l}{$^{\mathrm{c}}$DIP reference sensor configuration with 6 sensors} \\
        \multicolumn{4}{l}{$^{\mathrm{d}}$best sensor configuration $M_4(0.5)$ in terms of mesh error} \\
        \multicolumn{4}{l}{$^{\mathrm{e}}$full basic sensor configuration with 25 sensors}
    \end{tabular}
\end{table}

Since the focus of our work is on the evaluation of different sensor configurations, we take the \gls{dnn}-model from~\cite{Huang2018} as is, and train it for each sensor subset.
The training procedure consists of 50 epochs of training and 20~epochs of fine-tuning to achieve state-of-the-art accuracy and limit training time.

We log the positional, angular, SIP- and mesh error as well as jitter for each sensor combination.
As positional and angular errors are also expressed in the mesh error, we only report the mesh error and therefore still allow for comparison to [4-7].
The SIP-error is the mean orientation error of a joint-subset, i.e., elbow and knee.
The jitter evaluates the jerk over a sequence of time frames and therefore estimates the smootheness of the estimation.
All metrics are defined as L2-Norms and we average each metric over all sequences of the test dataset. 

To evaluate the accuracy-resource trade-off, we propose a combined metric~$M_i$ given in~\autoref{eq:combined_metric}, where $\lambda$ is a weight for hardware resources in relation to accuracy $e_i$, e.g. mesh error.
For assessment of hardware resources, we approximate costs like energy consumption, bandwidth, latency or wiring of a real sensor setup, by the number of sensors $i$.

\begin{equation}
    M_i(\lambda) = e_i \cdot (1 - \lambda) + \lambda \cdot i
    = e_i + \lambda \cdot (i - e_i) \label{eq:combined_metric}
\end{equation}

$\lambda$ is a design parameter and has to be a value between $0$ and \SI{100}{\percent}, where \SI{0}{\percent} corresponds to a system with a high emphasis on accuracy and \SI{100}{\percent} to a high emphasis on hardware resources, respectively.
$e_i$ has to be given in the same order of magnitude as the number of sensors to ensure comparability and avoid an additional scaling factor, i.e., mesh error has to be given in cm without unit.
The optimal configuration corresponds to the configuration with minimal $M_i(\lambda)$.

%% file: sections/4_results.tex
\section{Results and Discussion}

\begin{table}[tb]
    \centering
    \caption{\textbf{Best sensor configurations} identified for mesh error. The error range is given for the five best configurations. For two sensors, all possible configurations are listed.}
    \label{tab:best_configs_mesh}
    \begin{tabular}{c|c|c}
        \textbf{Number} &  & \textbf{Error range} \\
        \textbf{Sensors} & \textbf{Best configurations} & \textbf{(cm)} \\ \hline
         2 & [2, 0], [7, 0], [1, 0], [20, 0] & 12.80 - 13.50 \\ \hline
         3 & [16, 17, 0], [21, 22, 0], [12, 13, 0], &\\
         & [10, 11, 0], [5, 6, 0] & 7.55 - 9.11\\ \hline
         4 & [2, 16, 17, 0], [7, 16, 17, 0], &\\
         & [1, 16, 17, 0], [7, 21, 22, 0], &\\
         & [1, 21, 22, 0] & 6.03 - 6.66 \\ \hline
         5 & [5, 6, 21, 22, 0], &\\
         & [5, 6, 16, 17, 0], &\\
         & [2, 20, 16, 17, 0], &\\
         & [1, 20, 16, 17, 0], &\\
         & [2, 20, 21, 22, 0] & 5.59 - 5.79 \\ \hline
         6 & [7, 12, 13, 21, 22, 0], &\\
         & [2, 10, 11, 21, 22, 0], &\\
         & [7, 10, 11, 21, 22, 0], &\\
         & [20, 5, 6, 16, 17, 0], &\\
         & [7, 16, 17, 18, 19, 0] & 4.92 - 5.01 \\ \hline
         7 & [2, 20, 12, 13, 21, 22, 0], &\\
         & [2, 20, 10, 11, 16, 17, 0], &\\
         & [7, 20, 12, 13, 21, 22, 0], &\\
         & [7, 20, 16, 17, 23, 24, 0], &\\
         & [2, 20, 10, 11, 21, 22, 0] & 4.40 - 4.45 \\ \hline
         8 & [2, 10, 11, 18, 19, 21, 22, 0], &\\
         & [2, 10, 11, 16, 17, 23, 24, 0], &\\
         & [2, 12, 13, 21, 22, 23, 24, 0], &\\
         & [20, 5, 6, 18, 19, 21, 22, 0], &\\
         & [2, 10, 11, 14, 15, 16, 17, 0] & 3.80 - 3.90 \\ \hline
         9 & [2, 20, 12, 13, 18, 19, 21, 22, 0],  &\\
         & [7, 20, 12, 13, 18, 19, 21, 22, 0],  &\\
         & [2, 20, 10, 11, 18, 19, 21, 22, 0],  &\\
         & [2, 20, 12, 13, 21, 22, 23, 24, 0],  &\\
         & [7, 20, 10, 11, 18, 19, 21, 22, 0] & 3.22 - 3.28 \\ \hline
         10 & [1, 2, 20, 3, 4, 12, 13, 21, 22, 0], &\\
         & [2, 5, 6, 12, 13, 18, 19, 21, 22, 0], &\\
         & [1, 5, 6, 10, 11, 21, 22, 23, 24, 0], &\\
         & [1, 5, 6, 12, 13, 18, 19, 21, 22, 0], &\\
         & [2, 5, 6, 12, 13, 21, 22, 23, 24, 0] & 3.21 - 3.60 \\
    \end{tabular}
\end{table}

\begin{figure*}
    \centering
    \begin{tabular}{ c c c }
    \includegraphics[width=0.3\textwidth]{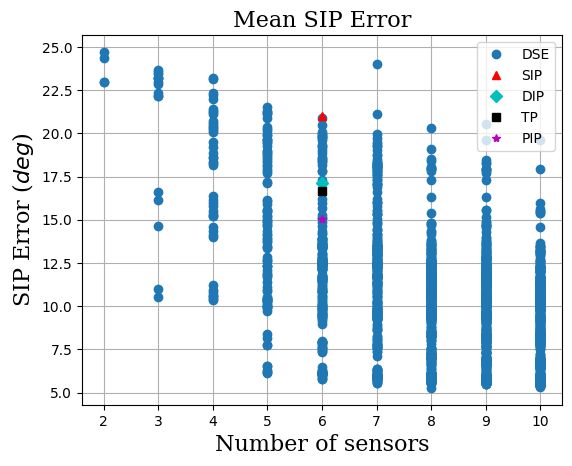}
        \label{fig:num_sip} \hfill &
    \includegraphics[width=0.3\textwidth]{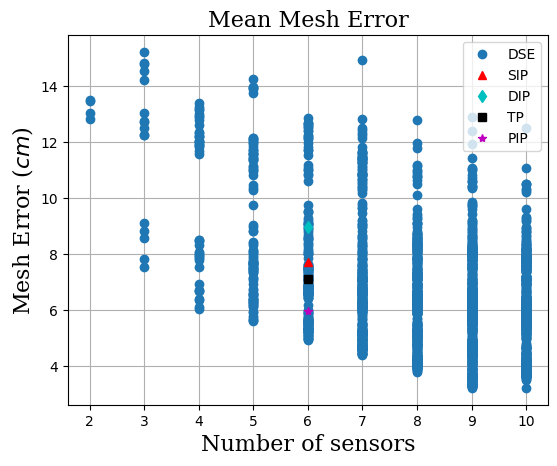}
        \label{fig:num_mesh} \hfill &
    \includegraphics[width=0.3\textwidth]{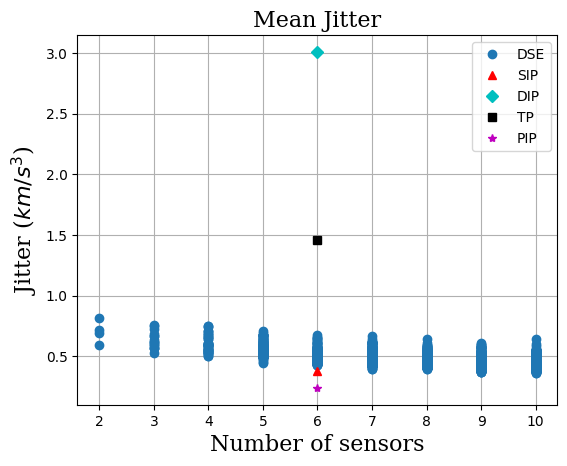}
        \label{fig:num_jitter} \\
    \end{tabular}
   \caption{\textbf{Accuracy of the sensor configurations} for different numbers of sensors.}
   \label{fig:num_error}
\end{figure*}

\begin{figure*}
    \centering
    \begin{tabular}{ c c c }
        \includegraphics[width=0.3\textwidth]{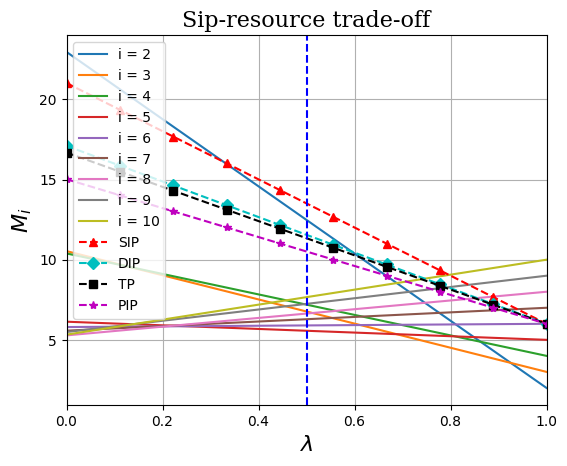} \hfill & \includegraphics[width=0.3\textwidth]{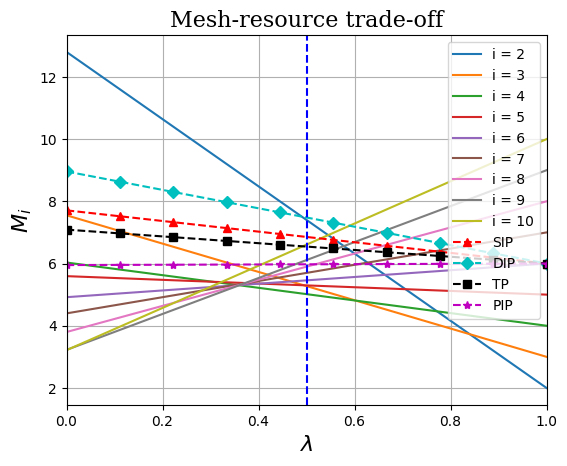} \hfill & \includegraphics[width=0.3\textwidth]{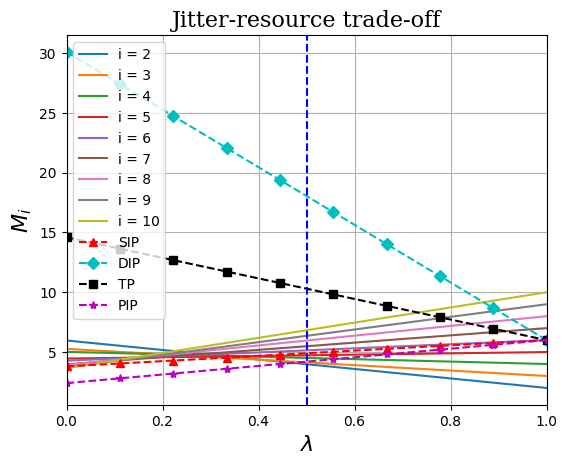} \\
        \small (a) $M_i(\lambda)$ for SIP-error in cm
        \label{fig:hw_sip} & 
        \small (b) $M_i(\lambda)$ for mesh in cm
        \label{fig:hw_mesh} &
        \small (c) $M_i(\lambda)$ for jitter in $0.1\frac{km}{s^3}$
        \label{fig:hw_jitter}
    \end{tabular}
    \caption{\textbf{Combined metric $M_i$} illustrating the accuracy-resource trade-off on varying hardware weight $\lambda$ for most accurate sensor configuration of $i$ sensors. The jitter is scaled by 0.1 to ensure the error metric being at the same order of magnitude as the number of sensors. The vertical blue line indicates a design with equal weight on prediction performance and hardware costs ($\lambda = \SI{50}{\percent}$).}
    \label{fig:hw_error}
\end{figure*}

For analysis of our methodology, we first describe some general findings and then go into detail of the \acrfull{dse}.


As shown in~\autoref{tab:sota_dl_hpe_evaluation}, due to additional white noise in the synthetic \gls{imu}-data, we improve the SIP and mesh error compared to the \gls{dip}~\cite{Huang2018} reference configuration by \SI{9.3}{\percent} and \SI{34.4}{\percent}, respectively.
The smoothed synthesis with $n=4$ frames reduces the jitter by \SI{83.7}{\percent}.
Compared to \cite{Huang2018}, our full basic configuration of all 25 sensors improves the SIP error by \SI{83.8}{\percent}, mesh error by \SI{78.4}{\percent} and jitter by \SI{84.4}{\percent}, compared to \cite{Xinyu2022}, only jitter is degraded.


However, the high number of sensors in the basic configuration does not consider any hardware-awareness.
Therefore, we analyze the error metrics results of our \gls{dse} depicted in~\autoref{fig:num_error}.
For SIP- and mesh error, many sensor configurations are more accurate than the state-of-the-art implementations from~[4-7].
The jitter matches the scale of the two-step methods, which optimize their first prediction with kinematic constraints like \gls{sip}, TransPose and \gls{pip}.
This originates from the additional white noise in our smoothed synthetic data.

For each error metric, we present the five best sensor configurations in~\autoref{tab:best_configs_mesh}, and count the occurrence of each sensor in the five best and worst setups, see~\autoref{tab:occ_5_best_worst}.

The sensor pair with ID 21 and 22, i.e., wrist, occurs the most frequently in the most accurate configurations and rarely appears in inaccurate configurations.
Consequently, the wrists are important positions to place sensors.
Similarly, the elbow sensors with ID 12 and 13 and the head sensor with ID 20 are included in top-ranked configurations and therefore contribute to an accurate \acrfull{hpe}.
In contrast, the back sensors with ID 1 and 7 are the most often part of inaccurate configurations, which is most likely because of their redundancy to the root node and the higher occurrence in setups with a low number of sensors with no limb sensor available.
Consequently, for an envisaged low number of sensors, positions at the end of the extremities should be preferred over an alignment at the back.

The foot sensors with ID 23 and 24 contribute to many inaccurate configurations.
In summary, sensors mounted on the upper body can be found much more often in accurate configurations and sensors mounted at the lower limbs contribute more often to inaccurate configurations.
This implies a higher importance of upper-body sensing than lower-limb sensing for an accurate full-body \gls{hpe}.
However, we are not sure if this implies a general rule or results from imbalanced data, which may be influenced by the dominance of upper-body motions in the dataset.
To finally conclude on the importance of upper or lower body sensing, a detailed analysis of the learning dataset and the focus of the exact application is needed.
In our exemplary use case of rehabilitation exercises for stroke patients, the restriction to focus sensing on one body-half would depend on the actual exercises and patients.

For a more precise analysis of the accuracy-resource trade-off, we show our combined metric $M_i(\lambda)$ from~\autoref{eq:combined_metric} for each best sensor configuration with 2 to 10 sensors separately for SIP-, mesh error and jitter in~\autoref{fig:hw_error}.
For a low importance of hardware costs $\lambda~\rightarrow~\SI{0}{\percent}$, $M_i$ equals exactly the \gls{hpe} accuracy and for high $\lambda \rightarrow \SI{100}{\percent}$, $M_i$ is defined by the number of sensors, respectively.
In between, the beneficial sensor configuration depends on the relative importance of accuracy and resources, which has to be defined by a system designer respecting the specific requirements of the application.

\begin{figure}[tb]
    \centering
    \includegraphics[scale=0.85]{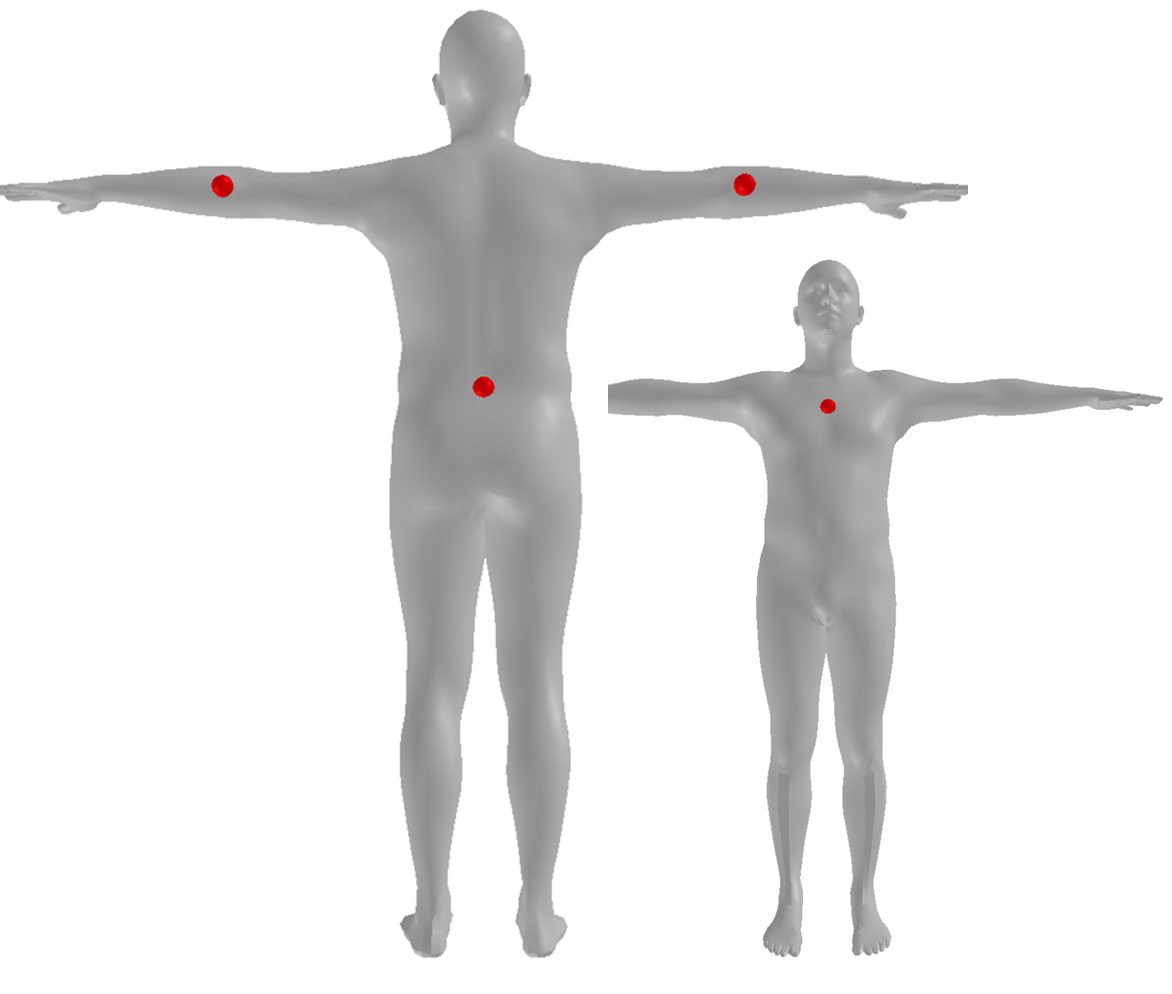}
    \caption{\textbf{Resulting optimal sensor configuration} for equal importance of accuracy and related hardware effort.}
    \label{fig:dsebest4sens}
\end{figure}

In case of an equal importance of accuracy and number of sensors ($\lambda = \SI{50}{\percent}$), all our configurations achieve a better SIP-resource trade-off than state of the art.
We identify five sensors as beneficial for SIP-, four sensors for mesh- and two sensors for the jitter-resource trade-off, respectively.
As the mesh error captures the human body in total, we focus on this error for a final decision and consider the sensor configuration with sensor IDs 0, 2, 16 and 17 (\autoref{fig:dsebest4sens}) as optimal to fulfill the accuracy-resource trade-off with equal importance of \gls{hpe} accuracy and related hardware effort in real applications.
This configuration achieves a mesh error of~\SI{6.03}{\cm} reducing the reference of~\gls{dip} by~\SI{32.7}{\percent} and increasing \gls{pip} by~\SI{1.3}{\percent}, while using two sensors less.

\begin{table}[tb]
    \centering
    \caption{\textbf{Occurrences of a sensor in the five best/worst configurations} per sensor ID and metric. The first and second value represent the number of occurrences in the best and worst configurations, respectively. The root sensor with ID 0 is present in each configuration. Due to symmetry, some sensors occur in pairs. Accurate configurations are highlighted in bold and inaccurate configurations with a gray background, respectively.}
    \label{tab:occ_5_best_worst}
    \begin{tabular}{c|c|c|c|c}
        \textbf{ID} & \textbf{SIP} & \textbf{Mesh} & \textbf{Jitter} & \textbf{Limb location} \\ \hline
        0 & 44 & 44 & 44 & Pelvis \\
        1 & \textbf{14} / \colorbox{gray!50}{15} & 7 / \colorbox{gray!50}{16} & 5 / \colorbox{gray!50}{32} & Back \\
        2 & 5 / 9 & \textbf{18} / 3 & 7 / \colorbox{gray!50}{19} & Sternum \\
        3, 4 & 5 / 4 & 1 / 7 & 4 / 16 & Thigh \\
        5, 6 & 12 / 2 & 9 / 1 & 14 / 5 & Shoulder \\
        7 & 7 / \colorbox{gray!50}{26} & 10 / \colorbox{gray!50}{20} & 6 / \colorbox{gray!50}{27} & Upper back \\
        8, 9 & 11 / 10 & 0 / 13 & 0 / 9 & Thigh \\
        10, 11 & \textbf{15} / 1 & 11 / 1 & 5 / 1 & Upper arm \\
        12, 13 & \textbf{22} / 14 & 12 / 14 & \textbf{21} / 4 & Upper arm \\
        14, 15 & \textbf{14} / 10 & 1 / 11 & 9 / 6 & Thigh \\
        16, 17 & 4 / 9 & 13 / 11 & \textbf{15} / 3 & Elbow \\
        18, 19 & 0 / 12 & 9 / 10 & 13 / 11 & Shin \\
        20 & 12 / 14 & \textbf{17} / 7 & \textbf{16} / 10 & Head \\
        21, 22 & 10 / 1 & \textbf{24} / 2 & 7 / 3 & Wrist \\
        23, 24 & 0 / \colorbox{gray!50}{17} & 6 / \colorbox{gray!50}{19} & 7 / 10 & Ankle \\
    \end{tabular}
\end{table}

%% file: sections/5_conclusion.tex
\section{Conclusion}
In summary, we developed a methodology to perform \acrfull{dse} for \acrfull{hpe} from sparse \gls{imu}-sensing with regard to both prediction performance and hardware costs in real systems and exemplary showed its application in rehabilitation.
We therefore synthesized \gls{imu}-data from a body model dataset for different sensor setups, trained a deep learning model with this data and evaluated over 2,000 different configurations.
Within this \gls{dse}, we observed more accurate \glspl{hpe} with sensors placed on the upper body than on the lower limbs.
In applications with a favored low number of sensors, sensors should rather be placed on the limbs than in line at the back.
For assessment of the important accuracy-resource trade-off, we proposed a combined metric with variable importance of both prediction accuracy and hardware resources defined by system requirements and usable by engineers to improve the evaluation of their system.
Applying this metric, we identified a sensor network of four sensors at the pelvis, sternum and elbows as beneficial for a system with equal importance of accuracy and resources, resulting in a mesh error of~\SI{6.03}{\cm}, which improves state of the art by \SI{32.7}{\percent} and reduces the hardware effort from six to four sensors.
In the future, we will improve our methodology to support the development of real motion tracking systems with fabric-integrated \gls{imu}-sensors and attention to data privacy and resource-awareness in diverse health applications.